\documentclass[conference]{IEEEtran}
\IEEEoverridecommandlockouts
\usepackage{cite}
\usepackage{amsmath,amssymb,amsfonts}
\usepackage{algorithmic}
\usepackage{graphicx}
\usepackage{textcomp}
\usepackage{xcolor}
\def\BibTeX{{\rm B\kern-.05em{\sc i\kern-.025em b}\kern-.08em
    T\kern-.1667em\lower.7ex\hbox{E}\kern-.125emX}}
\begin{document}

\title{Cooperative Localization Utilizing Reinforcement Learning for 5G Networks \\
}

\author{\IEEEauthorblockN{1\textsuperscript{st} Ghazaleh Kia}
\IEEEauthorblockA{\textit{dept. of Computer Science} \\
\textit{University of Helsinki}\\
Helsinki, Finland \\
ghazaleh.kia@helsinki.fi}
\and
\IEEEauthorblockN{2\textsuperscript{nd} Laura Ruotsalainen}
\IEEEauthorblockA{\textit{dept. of Computer Science} \\
\textit{University of Helsinki}\\
Helsinki, Finland \\
laura.ruotsalainen@helsinki.fi}
}
\maketitle

\begin{abstract}
The demand for accurate localization has risen in recent years to enable the emerging of autonomous vehicles. To have these vehicles in the traffic ecosystem of smart cities, the need for an accurate positioning system is emphasized. To realize accurate positioning, collaborative localization plays an important role. This type of localization computes range measurements between vehicles and improves the accuracy of position by correcting the possibly faulty values of one of them by using the more accurate values of the other. 5G signals with the technology of Millimeter Wave (mmWave) support precise range measurements and 5G networks provide Device to Device (D2D) communication which improves collaborative localization.
The aim of this paper is to provide an accurate collaborative positioning for autonomous vehicles, which is less prone to errors utilizing reinforcement learning technique for selecting the most accurate and suitable range measurement technique for the 5G signal.

\end{abstract}

\begin{IEEEkeywords}
5G, collaborative localization, D2D communications, mmWave, reinforcement learning
\end{IEEEkeywords}

\section{Introduction}

To provide safe transportation for everyone in a traffic ecosystem with unmanned vehicles, the accurate position of every object in the system is required. Radio-based localization is the most used positioning method. The most used navigation technology is the Global Navigation Satellite System (GNSS) localization. It works very well in open areas such as highways. However, in dense areas where tall buildings
exist and  signal blockage as well as multipath degrades the performance, GNSS-based localization \cite{GNSS-book}  is not feasible for autonomous vehicles \cite{5G-proposal}. Therefore, GNSS-based localization is usually complemented taking the advantage of collaborative localization. To shed light on this issue, in some scenarios GNSS signals help to find the position of some vehicles. However, the other vehicles positions are not easily recognized. Therefore, a range measurement between vehicles will assist finding unknown vehicles positions using collaborative localization techniques. 

5G signals being characterized of different technologies are good solution for accurate range measurement and as a result, accurate localization. For avoiding collisions in urban areas, an accuracy of 10 cm is required for vehicles which is provided by 5G signals \cite{5G-proposal}. In fact, there are several main technologies characterizing 5G to make it a trustworthy option to provide communications which are beneficial for positioning purposes.

In this paper, the 5G technologies are introduced in the second section. The third section discusses opportunities provided by 5G technologies and explains the focus of this paper and the method to address the chosen challenge. Section four explains the presented method.

\section{5G Technologies}

The common technologies which are adopted in 5G networks and frequently discussed in recent localization research \cite{main2,main3,main4,main5} are as follows. The first technology is higher density base stations and access points, which is a result of device-centric architecture. This technology will provide required Line-of-Sight (LoS) links for mobile stations. The second technology is massive Multiple Input Multiple Output (MIMO). Massive MIMO as an extension of MIMO provides enhanced spectrum efficiency as well as better throughput and support accurate Angle of Arrival calculations. mmWave as the third technology supporting high bandwidth as well as high frequencies results in precise Time Difference of Arrival (TDoA) measurements. mmWave refers to the communications in the frequency of above 24GHz and is interesting for localization  \cite{two114}. D2D communication is another technology of 5G networks. This technology provides many LoS links between mobile terminals. These links can be used to extract direction or range measurement between devices and provide the required data to distribute the localization task \cite{main3}. All of these technologies provide new opportunities in the field of localization to improve the accuracy of position solutions.

\section{Collaborative Positioning in 5G Networks}

Considering the technologies characterizing 5G networks as well as the challenges introduced in \cite{main4}, there are three main opportunities in the research field of 5G-based localization. The first opportunity is accurate mmWave propagation modeling. Using 5G networks, mobile communications work with higher frequencies. Thus, scientists frequently use waveform, beams, and channel modeling to deal with this challenge \cite{main5}. The second challenge is efficient channel parameter estimation. To address this challenge, researchers take the advantage of signal processing techniques such as compressive sensing as well as estimation methods and follow common steps \cite{jukkaCS}. The third research opportunity is cooperative localization in 5G networks. 

Taking the advantage of reinforcement learning \cite{ML1, ML2} methods and utilizing 5G networks, our work focuses on the third opportunity of cooperative localization. In this work, a network of vehicles being connected to each other with the help of 5G signals and synchronized are considered. D2D links exist between vehicles organized with the mmWave signals. The range information can be extracted from these signals using different methods.
Method A) is Time of Arrival (ToA) \cite{ToA}. For this method, signal propagation delay is calculated to find the distance between the transmitter and the receiver. In this method, it is presumed that LoS signal is available for the computation of signal propagation delay. Besides, this method is highly influenced by multipath. Method B) is Time Difference of Arrival (TDoA) \cite{TDoA}. In TDoA, the time difference of two signals arriving from two different anchors is calculated. This method decreases potential errors. Method C) is Received Signal Strength (RSS) \cite{RSS}. In this method, the strength of the received signal represents the distance between the vehicles. Although this technique usually results in the least accurate values, it is less influenced by multipath. 

\begin{figure}[htbp]
\centerline{\includegraphics[width=7.7cm]{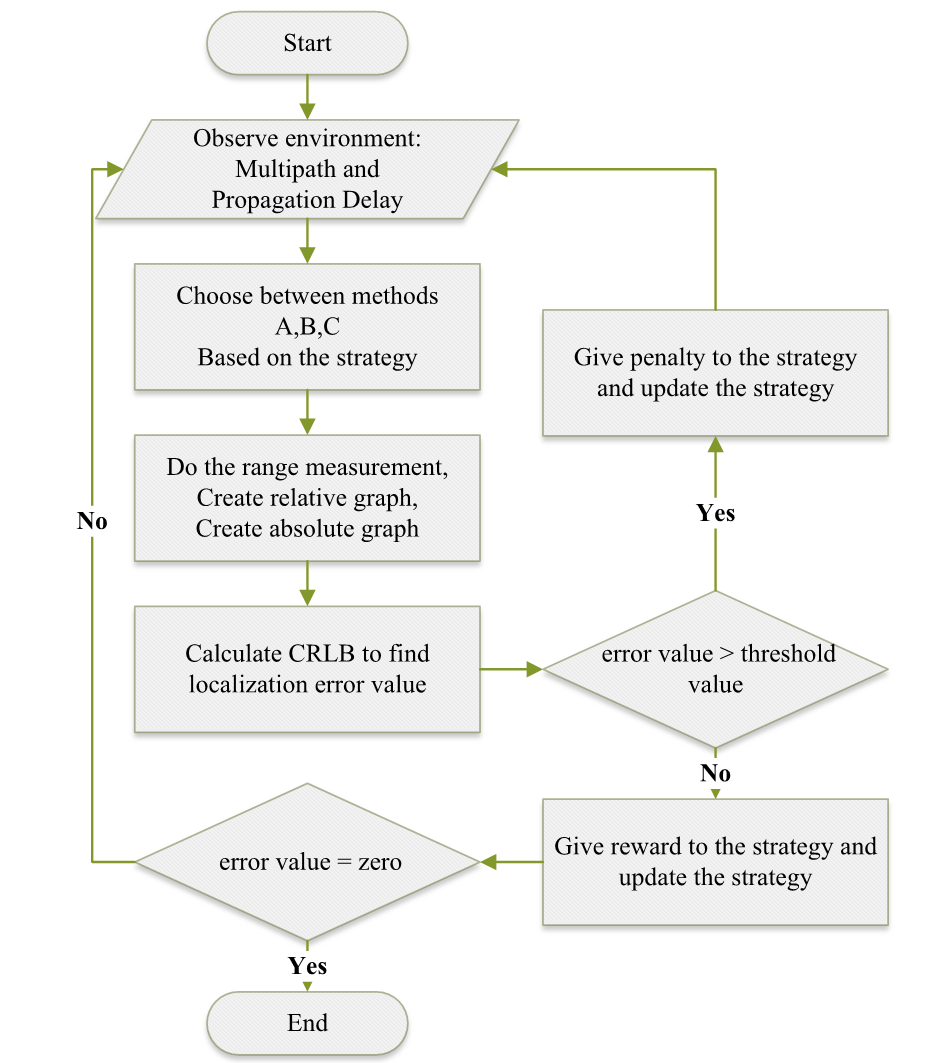}}
\caption{Reinforcement Learning in 5G Collaborative Localization.}
\label{fig}
\end{figure}

Each of these range measurement method has its benefit and drawback. In this paper, we provide a machine learning technique to learn from the results and decide to choose between these methods for each D2D signal range measurement.
After doing range measurements a relative graph of the mobile nodes will be created. Some nodes which have known locations, act as anchor nodes. Finally, a global transformation occurs to map the relative graph of the nodes (vehicles) to absolute graph \cite{main3}. This mapping is done with the help of anchor nodes position.
Then the localization will be evaluated using the minimum unbiased variance estimation. Cramer-Rao Lower Bound (CRLB) technique is frequently used for evaluation of the localization algorithm performance \cite{main2}. This evaluation can help improve the range measurement technique as shown in Fig.~\ref{fig} and explained in the next section.

\section{Using Reinforcement Learning to Improve Localization}
To make the system less prone to error and more accurate, we use reinforcement learning method to make the system choose the best range measurement technique to be used for each signal. Observing the environment, propagation delay of the signal using signal propagation model is available. Multipath is also available from the channel model. The system will first use one of the range measurement techniques based on a pre-defined strategy. Then it will do the range measurement using the chosen technique in previous step. Based on the range values, a relative graph is created and based on the anchor nodes location, a global transferring to absolute graph and localization is done. Then, an evaluation using CRLB occurs. By help of CRLB, the error of localization is calculated. If the error is less than a pre-defined threshold value, the strategy to choose from A,B,C methods receives a reward but if it is bigger than the threshold value, it receives a penalty. Thus, the strategy to choose range measurement method improves based on these penalties and rewards. Next time it will choose the range measurement technique based on the updated strategy and iterates until the absolute localization error equals zero.

 \bibliographystyle{./IEEEtran}
\bibliography{IEEEabrv,bibliography}

\begin{thebibliography}{10}
\providecommand{\url}[1]{#1}
\csname url@samestyle\endcsname
\providecommand{\newblock}{\relax}
\providecommand{\bibinfo}[2]{#2}
\providecommand{\BIBentrySTDinterwordspacing}{\spaceskip=0pt\relax}
\providecommand{\BIBentryALTinterwordstretchfactor}{4}
\providecommand{\BIBentryALTinterwordspacing}{\spaceskip=\fontdimen2\font plus
\BIBentryALTinterwordstretchfactor\fontdimen3\font minus
  \fontdimen4\font\relax}
\providecommand{\BIBforeignlanguage}[2]{{%
\expandafter\ifx\csname l@#1\endcsname\relax
\typeout{** WARNING: IEEEtran.bst: No hyphenation pattern has been}%
\typeout{** loaded for the language `#1'. Using the pattern for}%
\typeout{** the default language instead.}%
\else
\language=\csname l@#1\endcsname
\fi
#2}}
\providecommand{\BIBdecl}{\relax}
\BIBdecl

\bibitem{GNSS-book}
P.~Groves, \emph{Principles of GNSS, Inertial, and Multisensor Integrated
  Navigation Systems, Second Edition}, 03 2013.

\bibitem{5G-proposal}
K.~Witrisal \emph{et~al.}, ``Whitepaper on new localization methods for 5g
  wireless systems and the internet-of-things,'' 04 2018.

\bibitem{main2}
R.~M. {Buehrer}, H.~{Wymeersch}, and R.~M. {Vaghefi}, ``Collaborative sensor
  network localization: Algorithms and practical issues,'' \emph{Proceedings of
  the IEEE}, vol. 106, no.~6, pp. 1089--1114, June 2018.

\bibitem{main3}
P.~Zhang, J.~Lu, Y.~Wang, and Q.~Wang, ``Cooperative localization in 5g
  networks: A survey,'' \emph{ICT Express}, vol.~3, 03 2017.

\bibitem{main4}
F.~Wen, H.~Wymeersch \emph{et~al.}, ``A survey on 5g massive mimo
  localization,'' \emph{Digital Signal Processing}, vol.~94, 05 2019.

\bibitem{main5}
A.~{Shahmansoori}, G.~E. {Garcia}, G.~{Destino}, G.~{Seco-Granados}, and
  H.~{Wymeersch}, ``Position and orientation estimation through millimeter-wave
  mimo in 5g systems,'' \emph{IEEE Transactions on Wireless Communications},
  vol.~17, no.~3, pp. 1822--1835, March 2018.

\bibitem{two114}
H.~{Wymeersch} \emph{et~al.}, ``5g mmwave positioning for vehicular networks,''
  \emph{IEEE Wireless Communications}, vol.~24, no.~6, pp. 80--86, Dec 2017.

\bibitem{jukkaCS}
J.~{Talvitie}, M.~{Koivisto}, T.~{Levanen}, M.~{Valkama}, G.~{Destino}, and
  H.~{Wymeersch}, ``High-accuracy joint position and orientation estimation in
  sparse 5g mmwave channel,'' in \emph{ICC 2019 - 2019 IEEE International
  Conference on Communications (ICC)}, May 2019, pp. 1--7.

\bibitem{ML1}
S.~Pandey, ``Localization adopting machine learning techniques in wireless
  sensor networks,'' \emph{International Journal of Computer Sciences and
  Engineering}, vol.~6, pp. 366--374, 01 2018.

\bibitem{ML2}
T.~Le and S.~Moh, ``Reinforcement-learning-based topology control for wireless
  sensor networks,'' 12 2016, pp. 22--27.

\bibitem{ToA}
X.~{Cui}, T.~A. {Gulliver}, J.~{Li}, and H.~{Zhang}, ``Vehicle positioning
  using 5g millimeter-wave systems,'' \emph{IEEE Access}, vol.~4, pp.
  6964--6973, 2016.

\bibitem{TDoA}
G.~Mao and B.~Fidan, \emph{Localization Algorithms and Strategies for Wireless
  Sensor Networks}, 12 2009.

\bibitem{RSS}
M.~Khan, N.~Saeed, A.~Ahmad, and C.~Lee, ``Location awareness in 5g networks
  using rss measurements for public safety applications,'' \emph{IEEE Access},
  vol.~PP, pp. 1--1, 09 2017.

\end{thebibliography}

\end{document}